# Teaching Leśniewski's Prothetic with a Natural Deduction System

Pierre Joray

**Abstract**
Protothetic is one of the most stimulating systems for propositional logic. Including quantifiers and an inference rule for definitions, it is a very interesting mean for the study of many questions of metalogic. Unfortunately, it only exists in an axiomatic version, far too complicated and unusual to be easily understood by nowadays students in logic. In this paper, we present a system which is a natural deduction (in Fitch-Jaśkowski's style) version of protothetic. According to us, this system is adequate for teaching Leśniewski's logic to students accustomed to natural deduction.

## 1. What is protothetic?

Leśniewski's protothetic is one of the richest systems of propositional logic[1]. It is often described as a "generalized" system for the propositional calculus. This means that it includes not only the standard connectives, but also the use of quantifiers binding propositional variables, variables for connectives and in fact variables for functions of any order or category (like connectives, properties of connectives, operations on connectives, properties of such operations, etc). One of the important features of such an ambitious calculus is that its formal language cannot be restricted once and for all to any set of categories. On the contrary, it must always remain open to the infinitely many possibilities of new and more complex categories to be further introduced. Conceiving protothetic, Leśniewski faced with two main difficulties for which he discovered very interesting solutions.

The first difficulty is that the notion of *well formed formula* had to be relativised to a stage or a development of the formal language. Leśniewski's solution is known as the *contextual* conception of formal language. This conception is grounded on the key-notions of what is called today a categorical grammar. The most famous development of this idea is K. Ajdukiewicz's notion of syntactic connexion (Ajdukiewicz 1935).

The second difficulty was the elaboration of a procedure for the introduction of new constants not only of already available categories, but also of new ones. Leśniewski's solution was to elaborate an explicitly codified procedure for definitions conceived as an inference rule of his system. Contrary to the standard view (originated in the *Principia Mathematica*), Leśniewski's definitions do not simply introduce convenient abbreviations, but make the language richer and play an important role in proofs.

Protothetic is not a system in the usual sense of this word, but a logical formal machinery which allows several developments. A development is comparable to a chain of definitional expansions in the standard paradigm. The main difference lies in the fact that a development in protothetic is not a plurality of systems with relations being studied from the metalanguage. All the stages of a development are grounded on a unique formal axiom basis, which warrants all along the development certain metalogical properties, like soundness and consistency.

Searching adequate axioms for protothetic, Leśniewski tried to have only one primitive connective. He made the choice of biconditional (material equivalence), for it allows to state definitions in a very natural form. But he was held up in doing this at the beginning by his inability to reduce all the other connectives in terms of quantified biconditionals. The solution was found by the young Alfred Tarski in his PhD dissertation (1923), by quantifying not just propositional variables, but also variables for unary connectives.

---

[1] For detailed presentations, see Słupecki (1953), Miéville (1984, 2001) and Urbaniak (2014).

If we adopt Leśniewski's notation for (universal) quantifiers, with which a formula like $(\forall p)(\forall q)((p \equiv q) \equiv (q \equiv p))$ is written $\lfloor pq \rfloor \lceil (p \equiv q) \equiv (q \equiv p) \rceil$, then a usual axiom basis for protothetic (cf. Słupecki 1954) is given by the following four axioms (in which quantifiers bind propositional variables and variables for unary connectives):

Ax1: $\lfloor pqr \rfloor \lceil (p \equiv q) \equiv ((r \equiv q) \equiv (p \equiv r)) \rceil$

Ax2: $\lfloor pq \rfloor \lceil (p \equiv q) \equiv \lfloor f \rfloor \lceil f(p) \equiv f(q) \rceil \rceil$

Ax3: $\lfloor pq \rfloor \lceil (p \equiv q) \equiv \lfloor f \rfloor \lceil (f(p) \equiv f(q)) \equiv (p \equiv q) \rceil \rceil$

Ax4: $\lfloor f \rfloor \lceil f(\lfloor p \rfloor \lceil p \rceil) \equiv (f(\lfloor p \rfloor \lceil p \rceil) \equiv \lfloor p \rfloor \lceil p \rceil) \equiv \lfloor q \rfloor \lceil f(\lfloor p \rfloor \lceil p \rceil) \equiv f(q) \rceil) \rceil$

and the following five inference rules:

- Rule of substitution (elimination of the universal quantifier);
- Rule of detachment (*modus ponens* for biconditional);
- Rule of distribution (distribution of the quantifier to both arguments of a biconditional);
- Rule of definition (for stating quantified biconditional as definition-thesis);
- Rule of extensionality (for stating thesis similar to Ax2, for any defined category).

## 2. Teaching protothetic

Leśniewski's systems have been strongly overshadowed by Hilbertian style of formal systems and by model theory. But the interest of the former for the metatheory of logic and for the philosophy of logic remains very large. Among the important topics which are concerned, one can mention at least the following: the theory of categories (which is a syntactical version of simple type theory), the theory of definitions (which gives a good tool for the study of the importance of definitions in the foundational programs), the theory of many-orders quantifiers (which extract the theory of quantifiers from their objectual interpretation), a renewed notion of adequate set of primitives in axiom theories, etc.

But, as one can imagine when just looking at the above axiom system, teaching Leśniewski's logic is a big challenge. It has only been taken up by a very few logicians. Today's students are accustomed to streamlined methods of working with propositional logic. Protothetic must appear to them very difficult to understand and to work with. The reason is not simply a question of complexity. First, the style of presentation is specific to an old period of time, unknown to the students (in this respect, teaching Hilbert's or even Frege's original systems is also a challenge). But, it is known by testimonies that, in everyday's works, Leśniewski himself used a sort of natural deduction system for finding proofs. The idea was to make assumptions, then to examine their consequences and to collect the results into conditionals or biconditionals. This is just like our students of logic are accustomed. Jaśkowski's well-known natural deduction system has probably been inspired by these uncodified practices (Jaśkowski 1934). Unfortunately, no one decided to work on such a system of inference rules in the case of protothetic.

The main problems with a natural deduction system for protothetic is the status of definitions and the codification of rules which must consider the fact that the notion of *well formed formula* is contextual and modified each time a new definition is stated.

In the following sections, we present such a system, using semi-codified inference rules. Like original protothetic, our system, called PND is not a system in the usual sense, but a basis of rules with a few primitive constants (biconditional and universal quantifiers) allowing different developments. Our purpose is not to present PND in a rigorous manner, but to show, by the way of an example of development, how the system works. In our example, we will see how to define the propositional constants for truth, falsity and negation. After each definition, we will also show how specific rules for the use of the new constants can be derived. A rigorous presentation of the rules of PND and a proof that the addition of a rule for extensionality makes the system equivalent to protothetic will be presented in a forthcoming paper.

## 3. A natural deduction system for protothetic

A development in the natural deduction system PND is a unique proof, which contains sub-deductions elaborated by applying the basic inference rules, described by the following schemas:

**Rule** hyp

n
⋮
n+m

| $\alpha_1$ hyp
⋮
| $\alpha_m$ hyp

**Rule** rep

n | $\alpha$
⋯
| $\alpha$   n, rep

OR

n | $\alpha$
|
| $\alpha$   n, rep

**Rule** $\equiv e$

n | $\alpha \equiv \beta$
m | $\alpha$
⋯
| $\beta$   n, m, $\equiv e$

OR

n | $\alpha \equiv \beta$
m | $\beta$
⋯
| $\alpha$   n, m, $\equiv e$

**Rule** $\equiv i$

l | | $\alpha$
  | ⋮
m | | $\beta$
n | | $\beta$
  | ⋮
o | | $\alpha$
| $\alpha \equiv \beta$   l-m, n-o, $\equiv i$

**Rule** $\equiv ass(g)$

n | $\alpha \equiv (\beta \equiv \gamma)$
⋯
| $(\alpha \equiv \beta) \equiv \gamma$   n, $\equiv ass(g)$

**Rule** sub

m | $\lfloor v_1 \cdots v_{n-1} v_n \rfloor \lceil \alpha \rceil$
⋯
| $\lfloor v_1 \cdots v_{n-1} \rfloor \lceil \alpha(v_n / \delta) \rceil$   m, sub $(v_n / \delta)$

where
- $v_1 \cdots v_{n-1} v_n$ are *n* different variables;
- $\delta$ is a *wff* of the same category as $v_n$;
- $\alpha(v_n / \delta)$ is the result of the replacement in $\alpha$ of all free occurrences of $v_n$ by $\delta$;
- all the free occurrences of variables in $\delta$ remain free after the replacement;
- the main quantifier is deleted if it is empty after the operation.

**Rule** gen

m | | $\lfloor v_1 \cdots v_n \rfloor$
  | ⋮
o | | $\alpha$
| $\lfloor v_1 \cdots v_n \rfloor \lceil \alpha \rceil$   m-o, gen

where
- $v_1 \cdots v_n$ are *n* different variables;
- the sub-deduction does not depend on hypotheses;
- no rep in the sub-deduction with free occurrences of $v_1 \cdots v_n$.

**Rule** def

$$m\left| \lfloor v_1 \cdots v_n \rfloor \lceil \#(v_1 \cdots v_i) \cdots (v_j \cdots v_n) \equiv \alpha(v_1 \cdots v_n) \rceil \right. \quad \text{def}$$

where
- $v_1 \cdots v_n$ are $n$ different variables, if $n = 0$ the main quantifier is deleted;
- $\alpha(v_1 \cdots v_n)$, the *definiens*, is a formula of the category of propositions containing only symbols previously available in the development;
- $\#(v_1 \cdots v_i) \cdots (v_j \cdots v_n)$, the *definiendum*, is a formula of the category of propositions with only one occurrence of each symbol (apart from symbols for parentheses); it contains only one constant $\#$, which is a new symbol (not already used in the development);
- the *definiendum* and the *definiens* have the same free variables.

This account of the different basic rules is of course informal. A completely and rigorous presentation would have been much longer and much more precise in particular concerning the way to manage with the notion of *well formed formula of a certain given category*. This notion is difficult to precise, because it is relative to what has already been written in the development. In other words, the notion of a well formed formula is context dependent. But all the tools needed for that rigorous description are available in Leśniewski's work (1992).

Now, the aim of this paper is not to reach this rigorous account, but to give a sufficiently precise description in order to use the system in a quite intuitive way.

### 4. An example of development

In the following pages, we are going to present an example of development. As one will see in the first lines, each time a sub-deduction is constructed and the variables kept free in this sub-deduction, one can refer to it further as a *derived rule*. So during the development, the set of available inference rules increases. By convention, we will always write the new derived rules on the right of the development.

| | | | | |
|---|---|---|---|---|
| 0.1 | $\lfloor pq \rfloor$ | $p \equiv q$ | hyp | **Rule** $\equiv com$ |
| 0.2 | | $q$ | hyp | (derived from 0.1-0.8) |
| 0.3 | | $p \equiv q$ | 0.1, rep | $n\left| \alpha \equiv \beta \right.$ |
| 0.4 | | $p$ | 0.2, 0.3, $\equiv e$ | $\left| \beta \equiv \alpha \right.$    $n, \equiv com$ |
| 0.5 | | $p$ | hyp | |
| 0.6 | | $p \equiv q$ | 0.1, rep | |
| 0.7 | | $q$ | 0.5, 0.6, $\equiv e$ | |
| 0.8 | | $q \equiv p$ | 0.1-0.4, 0.5-0.7, $\equiv i$ | |
| 0.9 | | $q \equiv p$ | hyp | |
| 0.10 | | $p \equiv q$ | 0.9, $\equiv com$ | |
| 0.11 | | $(p \equiv q) \equiv (q \equiv p)$ | 0.1-0.8, 0.9-0.10, $\equiv i$ | |
| 1 | $\lfloor pq \rfloor \lceil (p \equiv q) \equiv (q \equiv p) \rceil$ | | 0.1-0.11, gen | |

It is worth noting that there are no theorems of the system itself, but only of a given development. All theorems are written on the main vertical stroke. Line 1 is here the first theorem of our development.

| | | | |
|---|---|---|---|
| 1.1 | $\lfloor p \rfloor$ | $p$ | hyp |
| 1.2 | | $p$ | 1.1, rep |
| 1.3 | | $p \equiv p$ | 1.1-1.2, 1.1-1.2, $\equiv i$ |
| 2 | $\lfloor p \rfloor \lceil p \equiv p \rceil$ | | 1.1-1.3, gen |

| 2.1 | ⌊pqr⌋ | p ≡ q | | hyp | | **Rule** ≡ syll |
|---|---|---|---|---|---|---|
| 2.2 | | q ≡ r | | hyp | | (derived from 2.1-2.13) |
| 2.3 | | | p | hyp | m | α ≡ β |
| 2.4 | | | p ≡ q | 2.1, rep | n | β ≡ γ |
| 2.5 | | | q | 2.3, 2.4, ≡ e | | α ≡ γ    m, n, ≡ syll |
| 2.6 | | | q ≡ r | 2.2, rep | | |
| 2.7 | | | r | 2.5, 2.6, ≡ e | | |
| 2.8 | | | r | hyp | | |
| 2.9 | | | q ≡ r | 2.2, rep | | |
| 2.10 | | | q | 2.8, 2.9, ≡ e | | |
| 2.11 | | | p ≡ q | 2.1, rep | | |
| 2.12 | | | p | 2.10, 2.11, ≡ e | | |
| 2.13 | | p ≡ r | | 2.3-2.7, 2.8-2.12, ≡ i | | |

This last part of the development has not the status of a proof, but only that of a deduction. It allows us to state a convenient derived rule. The following part of the development leads to the proof of what is known as Łukasiewicz's axiom for the complete biconditional calculus with quantifiers binding propositional variables (cf. theorem on line 4). This will be sufficient to show that our development is (potentially) at least as strong as this calculus.

| 2.14 | ⌊pqr⌋ | | p ≡ (q ≡ r) | hyp | | |
|---|---|---|---|---|---|---|
| 2.15 | | | (p ≡ q) ≡ r | 2.14, ≡ ass(g) | | |
| 2.16 | | | (p ≡ q) ≡ r | hyp | | **Rule** ≡ ass(d) |
| 2.17 | | | r ≡ (p ≡ q) | 2.16, ≡ com | | (derived from 2.16-2.23) |
| 2.18 | | | (p ≡ q) ≡ (q ≡ p) | 1, rep, sub ( p / p , q / q ) | n | (α ≡ β) ≡ γ |
| 2.19 | | | r ≡ (q ≡ p) | 2.17, 2.18, ≡ syll | | α ≡ (β ≡ γ)    n, ≡ ass(d) |
| 2.20 | | | (r ≡ q) ≡ p | 2.19, ≡ ass(g) | | |
| 2.21 | | | (q ≡ r) ≡ (r ≡ q) | 1, rep, sub ( p / q , q / r ) | | |
| 2.22 | | | (q ≡ r) ≡ p | 2.21, 2.21, ≡ syll | | |
| 2.23 | | | p ≡ (q ≡ r) | 2.22, ≡ com | | |
| 2.24 | | (p ≡ (q ≡ r)) ≡ ((p ≡ q) ≡ r) | | 2.14-2.15, 2.16-2.23, ≡ i | | |
| 3 | ⌊pqr⌋ | ⌈(p ≡ (q ≡ r)) ≡ ((p ≡ q) ≡ r)⌉ | | 2.14-2.24, gen | | |
| 3.1 | ⌊pq⌋ | p | | hyp | | **Rule** ≡ conj |
| 3.2 | | q | | hyp | | (derived from 3.1-3.7) |
| 3.3 | | | p | hyp | m | α |
| 3.4 | | | q | 3.2, rep | n | β |
| 3.5 | | | q | hyp | | α ≡ β    m, n, ≡ conj |
| 3.6 | | | p | 3.1, rep | | |
| 3.7 | | p ≡ q | | 3.3-3.4, 3.5-3.6, ≡ i | | |
| 3.8 | ⌊pqr⌋ | | (r ≡ q) ≡ (p ≡ r) | hyp | | |
| 3.9 | | | ((r ≡ q) ≡ p) ≡ r | 3.8, ≡ ass | | |
| 3.10 | | | (r ≡ (q ≡ p)) ≡ ((r ≡ q) ≡ p) | 3, rep, sub ( p / r , q / q , r / p ) | | |
| 3.11 | | | (r ≡ (q ≡ p)) ≡ r | 3.10, 3.9, ≡ syll | | |
| 3.12 | | | r ≡ (r ≡ (q ≡ p)) | 3.11, ≡ com | | |
| 3.13 | | | (r ≡ r) ≡ (q ≡ p) | 3.12, ≡ ass | | |
| 3.14 | | | r ≡ r | 2, rep, sub ( p / r ) | | |

| | | | |
|---|---|---|---|
| 3.15 | | $q \equiv p$ | 3.13, 3.14, $\equiv e$ |
| 3.16 | | $p \equiv q$ | 3.15, $\equiv com$ |
| 3.17 | | $p \equiv q$ | hyp |
| 3.18 | | $q \equiv p$ | 3.17, $\equiv com$ |
| 3.19 | | $r \equiv r$ | 2, rep, sub ($p/r$) |
| 3.20 | | $(r \equiv r) \equiv (q \equiv p)$ | 3.18, 3.19, $\equiv conj$ |
| 3.21 | | $r \equiv (r \equiv (q \equiv p))$ | 3.20, $\equiv ass$ |
| 3.22 | | $(r \equiv (q \equiv p)) \equiv ((r \equiv q) \equiv p)$ | 3, rep, sub ($p/r, q/q, r/p$) |
| 3.23 | | $r \equiv ((r \equiv q) \equiv p)$ | 3.21, 3.22, $\equiv syll$ |
| 3.24 | | $((r \equiv q) \equiv p) \equiv r$ | 3.23, $\equiv com$ |
| 3.25 | | $(r \equiv q) \equiv (p \equiv r)$ | 3.24, $\equiv ass$ |
| 3.26 | | $(p \equiv q) \equiv ((r \equiv q) \equiv (p \equiv r))$ | 3.8-3.16, 3.17-3.25, $\equiv i$ |
| 4 | $\lfloor pqr \rfloor \lceil (p \equiv q) \equiv ((r \equiv q) \equiv (p \equiv r)) \rceil$ | | 3.8-3.26, gen     [Łukasiewicz's axiom] |

Now, we will state the definitions of the two propositional constants for truth and falsity. In order to be available everywhere later in the development, the definitions have to be written as theorems, on the main stroke.

| | | | |
|---|---|---|---|
| 5 | $\top \equiv \lfloor p \rfloor \lceil p \equiv p \rceil$ | Def | |
| 6 | $\bot \equiv \lfloor p \rfloor \lceil p \rceil$ | Def | |
| 7 | $\top$ | 5, 2, $\equiv e$ | |
| 7.1 | $\lfloor q \rfloor \; \bot$ | hyp | **Rule** $\bot e$ |
| 7.2 | $\bot \equiv \lfloor p \rfloor \lceil p \rceil$ | 6, rep | (derived from 7.1-7.4) |
| 7.3 | $\lfloor p \rfloor \lceil p \rceil$ | 7.1, 7.2, $\equiv e$ | n $\mid \bot$ |
| 7.4 | $q$ | 7.3, sub ($p/q$) | $\alpha$   n, $\bot e$ |

In the following lines, we are going to show how the new constants can be used to mimic the biconditional's truth table by four theorems (lines 11-14).

| | | | |
|---|---|---|---|
| 8 | $\top \equiv \top$ | 2, sub ($p/\top$) |
| 9 | $\bot \equiv \bot$ | 2, sub ($p/\bot$) |
| 10 | $(\bot \equiv \top) \equiv (\top \equiv \bot)$ | 1, sub ($p/\bot, q/\top$) |
| 10.1 | $\bot$ | hyp |
| 10.2 | $\top \equiv \bot$ | 10.1, $\bot e$ |
| 10.3 | $\top \equiv \bot$ | hyp |
| 10.4 | $\top$ | 7, rep |
| 10.5 | $\bot$ | 10.3, 10.4, $\equiv e$ |
| 11 | $(\top \equiv \bot) \equiv \bot$ | 10.1-10.2, 10.3-10.5, $\equiv i$ |
| 12 | $(\bot \equiv \top) \equiv \bot$ | 10, 11, $\equiv syll$ |
| 13 | $(\top \equiv \top) \equiv \top$ | 7, 8, $\equiv conj$ |
| 14 | $(\bot \equiv \bot) \equiv \top$ | 7, 9, $\equiv conj$ |

The following lines are devoted to the definition of negation, and the derivation of its characteristic inference rules. Lines 18, 19 correspond to the truth table for negation.

| | | | |
|---|---|---|---|
| 15 | $\lfloor p \rfloor \lceil \sim p \equiv (p \equiv \bot) \rceil$ | Def |
| 16 | $\sim \top \equiv (\top \equiv \bot)$ | 15, sub ($p/\top$) |
| 17 | $\sim \bot \equiv (\bot \equiv \bot)$ | 15, sub ($p/\bot$) |
| 18 | $\sim \top \equiv \bot$ | 16, 11, $\equiv syll$ |

| | | | | |
|---|---|---|---|---|
| 19 | ~⊥≡⊤ | 17, 14, ≡ syll | | |
| 20 | ~⊥ | 19, 7, ≡ e | | |
| 20.1 | ⌊p⌋  p | hyp | **Rule** ⊥i | |
| 20.2 | ~p | hyp | (derived from 20.1-20.5) | |
| 20.3 | ~p ≡ (p ≡⊥) | 15, rep, sub (p / p) | m | α |
| 20.4 | p ≡⊥ | 20.2, 20.3, ≡ e | n | ~α |
| 20.5 | ⊥ | 20.1, 20.4, ≡ e | | ⊥  m, n, ⊥i |
| 20.6 | ⌊p⌋   p | hyp | | |
| 20.7 | ⊥≡⊥ | 9, rep | **Rule** ~~i | |
| 20.8 | p ≡ (⊥≡⊥) | 20.6, 20.7, ≡ conj | (derived from 20.6-20.13) | |
| 20.9 | (p ≡⊥) ≡⊥ | 20.8, ≡ ass | n | α |
| 20.10 | ~p ≡ (p ≡⊥) | 15, rep, sub (p / p) | | ~~α   n, ~~i |
| 20.11 | ~p ≡⊥ | 20.9, 20.10, ≡ syll | | |
| 20.12 | ~~p ≡ (~p ≡⊥) | 15, rep, sub (p/ ~p) | | |
| 20.13 | ~~p | 20.11, 20.12, ≡ e | | |
| 20.14 | ~~p | hyp | | |
| 20.15 | ~~p ≡ (~p ≡⊥) | 15, rep, sub (p/ ~p) | **Rule** ~~e | |
| 20.16 | ~p ≡⊥ | 20.14, 20.15, ≡ e | (derived from 20.14-20.22) | |
| 20.17 | ~p ≡ (p ≡⊥) | 15, rep, sub (p / p) | n | ~~α |
| 20.18 | (p ≡⊥) ≡ ~p | 20.17, ≡ com | | α     n, ~~e |
| 20.19 | (p ≡⊥) ≡⊥ | 20.16, 20.18, ≡ syll | | |
| 20.20 | p ≡ (⊥≡⊥) | 20.19, ≡ ass | | |
| 20.21 | ⊥≡⊥ | 9, rep | | |
| 20.22 | p | 20.20, 20.21, ≡ e | | |
| 20.23 | p ≡ ~~p | 20.6-20.13, 20.14-20.22, ≡ i | | |
| 21 | ⌊p⌋⌈p ≡ ~~p⌉ | 20.6-20.23, gen | | |

Unfortunately, the rule for the introduction of negation cannot be simply derived in the same manner as the previous ones. Like with all the rules requiring the construction of a sub-deduction, there is no explicit deduction with the general form of this rule. The following example of proof shows nevertheless that there is an available strategy corresponding to the introduction of negation:

| | | |
|---|---|---|
| 21.1 | ⌊pq⌋⌈p ≡ q⌉ | hyp |
| 21.2 | p ≡ ~p | 21.1, sub (p / p, q/ ~p) |
| 21.3 | ~p ≡ (p ≡⊥) | 15, rep, sub (p / p) |
| 21.4 | p ≡ (p ≡⊥) | 21.2, 21.3, ≡ syll |
| 21.5 | (p ≡ p) ≡⊥ | 21.4, ≡ ass |
| 21.5 | ⊥ | 21.5, 2, rep, sub (p / p), ≡ e |
| 21.6 | ⊥ | hyp |
| 21.7 | ⌈pq⌋⌈p ≡ q⌉ | 21.6, ⊥ e |
| 22 | ⌊pq⌋⌈p ≡ q⌉ ≡⊥ | 21.1-21.5, 21.6-21.7, ≡ i |
| 23 | ~⌊pq⌋⌈p ≡ q⌉ | 22, 15, sub (p /⌊pq⌋⌈p ≡ q⌉), ≡ e |

Here we could give a meta-rule matching this *reductio ad absurdum* strategy and stating that the existence of a sub-deduction beginning with a formula $\alpha$ as its single assumption and ending with ⊥ (here, lines 21.1-21.5) is sufficient to infer the formula $\sim\alpha$ (here, line 23).

## 5. Conclusion

It would have been too long here to show that PND is sufficiently strong to reach the complete classical propositional calculus. I can just say that that the following definition of the conjunction (due to Tarski, 1923) would have been adequate to do the job into my example of development:

24 $\quad \lfloor pq \rfloor \lceil (p \wedge q) \equiv \lfloor f \rfloor \lceil p \equiv (\lfloor r \rfloor \lceil p \equiv f(r) \rceil \equiv \lfloor r \rfloor \lceil q \equiv f(r) \rceil) \rceil \rceil \qquad$ Def

Nevertheless, the aim of this paper was just to show that a system very close to Fitch's and Jaśkowski's style of natural deduction systems is available to work with protothetic. With the exception of the rule for definitions, all the procedures and also the deductive strategies are very close to those to which today's students in logic are all accustomed.

Moreover, it is worth noting that the addition of two primitive rules for the introduction and elimination of Leśniewski's constant $\varepsilon$ would enable us to have a natural deduction system for Leśniewski's calculus of names (the system called *ontology*), which is a system containing standard predicate logic as one of his developments.

As a final note, it must be precised that PND is not equivalent to the full system of Leśniewski's protothetic, for it lacks the means for the proofs of what we can call the laws of extensionality. Laws of extensionality are the formulas that express for all categories what the following expresses just for the basic category of propositions:

$\lfloor pq \rfloor \lceil (p \equiv q) \equiv \lfloor f \rfloor \lceil f(p) \equiv f(q) \rceil \rceil$

According to me, this is not a defect of PND. On the contrary, a comparative study of PND and PND + Ext (the system obtained by the addition of Leśniewski's rule of extensionality) would open a quite stimulating research about the specificities of (still unknown) versions of protothetic open to intensional constants.

## References

Ajdukiewicz K. (1935). "Die syntaktische Konnexität", *Studia Philosophica* **1**, 1-27.

Jaśkowski S. (1934), "On the rules of suppositions in formal logic", Engl. tr. in *Studia Logica* **1**, 5-32.

Leśniewski S. (1992), *Collected Works*, Warsaw: Polish Scientific Publ. / Dordrecht: Kluwer.

Miéville D. (1984), *Un développement des systèmes logiques de Stanisław Leśniewski. Prothétique, ontologie, méréologie*, Bern: Peter Lang.

Miéville D. (2001), *Introduction à l'œuvre de S. Leśniewski, Fasc. I : la protothétique*. Hors série des *Travaux de logique*. Neuchâtel : Université.   (http://doc.rero.ch/record/208738?ln=fr).

Słupecki J. (1953), "St. Leśniewski's Protothetics", *Studia Logica* **1**, 44-112.

Tarski A. (1923), «Sur le terme primitif de la logistique», *Fundamenta Mathematicae* **4**, 196-220. Engl. tr. in Tarski A. *Logic, Semantics, Metamathematics*, Oxford: Clarendon, 1956.

Urbaniak R. (2014), *Leśniewski's Systems of Logic and Foundations of Mathematics*, Springer.